\documentstyle[prl,aps,amsmath,amssymb,multicol,epsf,epsfig]{revtex}
\begin{document}
\draft
\title{ The van der Waals Potential between Metastable Atoms and Solid
 Surfaces:\\ Novel Diffraction Experiments versus Theory}  
\author{R\"udiger~Br\"uhl, Peter~Fouquet, Robert~E.~Grisenti, and J.~Peter~Toennies}
\address{Max-Planck-Institut f\"ur Str\"omungsforschung, Bunsenstr. 10,
 37073 G\"ottingen, Germany} 
\author{Gerhard~C.~Hegerfeldt, Thorsten~K\"ohler, Martin~Stoll, and Christian~Walter}
\address{Institut f\"ur Theoretische Physik, Universit\"at G\"ottingen, 
 Bunsenstr. 9, 37073 G\"ottingen, Germany}
\maketitle
\begin{abstract}  
Highly polarizable metastable He* ($\mathrm{2^3S}$) and Ne*
($\mathrm{2^3P}$) atoms have been diffracted  from a 100\,nm period
silicon nitride transmission grating and 
the van der Waals coefficients $C_3$ for the interaction of the
excited atoms  with the silicon nitride surface have been determined from
the diffraction intensities out to the
10th order. The results agree with calculations based on the
non-retarded Lifshitz formula.
\end{abstract}

The van der Waals (vdW) force between atoms, molecules and solid
surfaces is of far reaching importance in many branches of physics,
chemistry, and biology \cite{Bruch97}. 
For larger
distances, retardation due to the exchange of virtual photons has to be
included,  while for distances much smaller than the smallest wavelength
a non-retarded approach can be used. The theoretical foundations for
atom-surface interactions were laid in the pioneering work of Lifshitz
\cite{Lifshitz56}. In this case the non-retarded vdW potential
 has the form $-C_3/l^3 $ in leading order,
where $l$ is the atom-surface separation and $C_3$ depends on the
atom, its electronic state, and on the  electronic states of the
solid. 

For {\it groundstate} rare gas atoms  the $C_3$ coefficients have
recently been  
measured with good accuracy \cite{Grisenti99}. Less is known about the van
der Waals interactions of  electronically excited  
{\it metastable} and {\it Rydberg atoms}, in particular the $C_3$
coefficient is not accurately known. 
Some time ago,  transmission
through narrow channels \cite{Anderson88} and  level shifts in
closed or semi-infinite cavities
\cite{Jhe87,Failache99} have been studied. Recently,
 inelastic electronic transitions 
on passage over a metal edge \cite{Boustimi01} and reflection from 
surfaces and (reflection) gratings \cite{Shimizu01} have been measured. 
Currently there  is great
interest in these potentials, in particular for
metastable helium which is widely used in atom optics \cite{Adams94} 
as well as in surface physics \cite{surf} and for which Bose-Einstein
condensation has recently been achieved \cite{Robert01}. The
atom-surface van der Waals potentials could soon become relevant in
guiding 
slow metastable rare gas atoms along microstructures
\cite{Sengstock01} or in studying collective effects of 
Bose-Einstein condensed  metastable He*  atoms in contact
with a surface. 

From a theoretical point of view atoms in excited states are of
particular interest. Their 
polarizability $\alpha$  is expected to increase as $n^7$, with
correspondingly 
much stronger interactions \cite{Ste83}. Therefore it is not obvious whether 
approximate formulae for the groundstate atom-surface vdW potential 
are still applicable for excited
atoms. Moreover, with the much stronger vdW interaction new effects
such as  higher multipole coefficients \cite{Hutson86} can be
expected.

In this article, an effective but simple experimental method
is used to determine the atom-surface vdW coefficient $C_3$ for
metastable rare gas atoms. It is based on diffraction of an atomic
beam from a nanostructured transmission grating with a period of only
100\,nm.
Modifications in the hierarchy of the intensities 
of the higher order maxima in the
diffraction pattern have been shown to be directly 
related to the strength $C_3$
of the atom-surface vdW potential. 
In these experiments only the
non-retarded regime is probed since the slit widths of about 50\,nm
are less than the distance at which retardation effects become significant. 

The main difference  to the case of rare gas groundstate atoms is the fact that
for the theoretical evaluation of the experimental data the very
convenient notion of an ``effective'' slit width can no longer be
employed, due to the much stronger surface interaction. Therefore  a
new approach is presented which can also be applied to other atoms and
molecules and materials \cite{Zeilinger2002}.

The diffraction apparatus has already been described in Ref.
\cite{Grisenti00}. The metastable 
atoms are produced by a discharge in the free-jet
expansion zone inside a sapphire nozzle
(aperture diameter $160\,\mu\rm{m}$) \cite{Fouquet98}. The He* in the
beam is 98\% in the $^3$S$_1$ state and 2\% in the $^1$S$_0$ state
\cite{Fouquet98} and the Ne* is 
85\% in the \,$^3$P$_2$ state and 15\%  in the $^3$P$_0$
state \cite{Siska93}. 
The stable operation of the
discharge limits the source pressure $P_0$ to the range $0.5-3\,\rm{bar}$, with
the consequence that the atomic beams have a rather broad velocity 
distribution  with $\Delta v / u \simeq 13.3\,\%$, where $\Delta v$ and $u$ denote
the full half width and the mean velocity, respectively.
The effective velocity spread could be reduced to about $\Delta
v/u\simeq 3\,\%$ by extracting the diffraction pattern for a given flight time
``slice'' through time-of-flight (TOF) spectra measured at 
closely spaced diffraction angles. 
After passing through the $0.72\,\rm{mm}$ diameter skimmer the beam is collimated
by two 5 mm high, 20 micron
(150 mm from the source) and 10 micron (1000 mm from the source) wide 
slits, before illuminating about $N=100$ slits of the silicon nitride transmission diffraction
grating with a calibrated period of 100\,nm, placed $150\,\rm{mm}$ behind the second collimation
slit.  
The depth $t = 53\,\mathrm{nm}$ and wedge angle $\beta = 11^\circ$
of the trapezoidally shaped grating bars were determined from 
transmission measurements \cite{Grisenti00} and the slit width at
infinite velocity was measured to be $s_0=66.8\,\mathrm{nm}$
\cite{Grisenti99}.  
Although the relative population of metastable atoms is only about 
10$^{-5}$ for both He* and Ne* the groundstate atoms are entirely
suppressed by the channel electron 
multiplier detector, mounted at a distance of $730\,\mathrm{mm}$
from the grating. 
The diffraction pattern is recorded by rotating the channeltron in
angular steps of about 0.1\,$\mu$rad with respect to an axis 
passing through the grating bars. 
The $25\,\mu\rm{m}$ wide detector slit, $r=430\,\rm{mm}$ downstream
from the grating, provides an angular resolution of 
$100\,\mu\rm{rad}$.   
TOF distributions were measured at each detector position for
5 (He*) and 8 minutes (Ne*). The peak shapes were fitted with a
Gaussian to determine the peak areas $I_n$, which were normalized to the
total over all peaks $I_{\mathrm{tot}}$. Fig. 1 shows 
typical angular distributions for He* at a time slice corresponding 
to 2347\,m/s (de Broglie wavelength $\lambda=42.5\,\mathrm{pm}$) and for Ne*
at 873\,m/s ($\lambda=22.6\,\mathrm{pm}$).
 
In the Fraunhofer limit $r\gg d$ the intensities of the $n$th order
maxima in the diffraction pattern are given by
\begin{equation}\label{intensity}
  I_n\propto\left |\frac{\sin(\frac{Nkd}{2}\sin\vartheta
      _n)}{\sin(\frac{kd}{2}\sin\vartheta _n)}
    f_{\mathrm{slit}}(\vartheta _n)\right |^2\mathrm ,
\end{equation}
where the atom momentum is
$mv=\hbar k=\frac{h}{\lambda}$ and the $n$th 
order diffraction angles are given by $\sin\vartheta _n=\frac{n\lambda}{d}$.
At high velocities (small $\lambda$) small anomalies in the observed
peak shapes at higher diffraction orders, which were attributed to
Fresnel effects, were shown to have no significant influence on the
interpolated Gaussian peak shapes.
 The slit
function $f_{\mathrm{slit}}(\vartheta)$ has the form\cite{Grisenti99}
\begin{equation}\label{slitamplitude}
f_{\mathrm{slit}}(\vartheta)=\frac{2\cos\vartheta}{\sqrt{\lambda}}
\int\limits_0^{s_0/2}{\mathrm
 d}\zeta\cos\left[k\sin\vartheta\left(\frac{s_0}{2}-\zeta\right)\right]\tau
(\zeta)\mathrm ,
\end{equation}
where the integration is over half the slit opening from the edge
($\zeta=0$) to the center ($\zeta=s_0/2$)
with $\zeta$ as impact parameter with respect to the
grating bar edge. In the
usual Eikonal approximation \cite{Joachain83} the amplitude
$\tau(\zeta)$ at different positions in the slit becomes 
\begin{equation}\label{tau}
  \tau(\zeta)={\mathrm exp}\left(-\frac{\mathrm i}{\hbar
      v}\int\limits_{-\infty}^{+\infty} {\mathrm d}z\,V_{\mathrm{att}}(z,\zeta)\right)\mathrm ,
\end{equation}
where  the $z$-axis is in the beam direction. The attractive potential
$V_{\mathrm{att}}$ is $-C_3/l^3$ for a plane. For a grating bar, due
to its finite extent, minor corrections occur which are taken into
account here.
Integration along a straight line trajectory for a given $\zeta$
yields, after some calculation,
\begin{equation}\label{tau2}
  \tau(\zeta)={\mathrm exp}\left(\frac{{\mathrm i}C_3t}{\hbar v\,\zeta
      ^3}\frac{1+\frac{t}{2\zeta}\tan\beta}
    {(1+\frac{t}{\zeta}\tan\beta)^2}\right)\,\mathrm .
\end{equation}
Because of the much larger value of $C_3$
the cumulant expansion of
Eq. (\ref{slitamplitude}) used in Ref. \cite{Grisenti99} was found not
to converge for metastables so that the 
convenient notion of an effective slit width and the 
formula of Ref. \cite{Grisenti99} for $I_n/I_{\mathrm{tot}}$ are no longer
applicable. Thus it was necessary to
calculate  $I_n/I_{\mathrm{tot}}$ from
Eqs.\,(\ref{intensity})-(\ref{tau2}) 
with $C_3$ as a parameter and use a least-square fit to
 the experimental values. This procedure is more
 complicated and more sensitive to experimental and numerical errors
 than that based on effective slit widths.
Unlike Ref.\,\cite{Grisenti99} the best fit was obtained without
a Debye-Waller damping factor  to
account for surface roughness \cite{Grisenti00}. 
The lack of sensitivity to these defects  for He* and
Ne* is attributed to the greater range of the potential, so that their
effect is smeared out.

The experimental results  $C_3$(He*)$=(4.1\pm 1.0)$ meV\,nm$^3$
and $C_3$(Ne*)$=(2.8\pm 1.0)$ meV\,nm$^3$ are, as expected, more than an order of
magnitude larger than the 
corresponding values  of
$C_3({\rm He})=(0.10 \pm 0.02)$ meV\,nm$^3$ 
and $C_3({\rm Ne})=(0.21\pm 0.04)$ meV\,nm$^3$ for the groundstate atoms
\cite{Grisenti99}.

Present approximations of atom-surface vdW forces are based on the
expression of Lifshitz \cite{Lifshitz56},
\begin{equation}\label{Lifshitz}
  C_3=\frac{\hbar}{4\pi}\int\limits _0^{\infty}{\mathrm d}\omega\,\alpha({\mathrm i}
  \omega)\,g({\mathrm i} \omega)\,\mathrm ,
\end{equation}
where $\alpha(\mathrm i\omega)$ is the dynamic polarizability of the
atom and
$g(\mathrm i\omega)$ is the corresponding response of the electrons of
the solid which is related to the dielectric function $\epsilon$ by
\begin{equation}\label{g_0}
  g({\mathrm i}\omega)= \frac{\epsilon ({\mathrm i}\omega)-1}{\epsilon ({\mathrm i}\omega)+1}\,\mathrm .
\end{equation}
 Unfortunately, $\alpha$ and $g$ are not known in general. Vidali and 
 Cole\cite{Vidali81} have studied  a model in which the partners are
 treated as single oscillators of  
 frequencies $E_{\mathrm a}/\hbar$ for the atom and $E_{\mathrm
S}/\hbar$ for the solid, i.e.
\begin{equation}\label{alpha}
  \alpha({\mathrm{i}}\omega) \approx
  \frac{\alpha(0)}{1+(\hbar\omega)^2/E_{\mathrm{a}}^2}
\end{equation}
and
\begin{equation}\label{g}
  g({\mathrm{i}}\omega) \approx
  \frac{g_0}{1+(\hbar\omega)^2/E_{\mathrm{S}}^2}\,\mathrm.
\end{equation}
It has been shown earlier by Tang \cite{Tang69} that the one-oscillator
approximation Eq.\,(\ref{alpha}) is correct to within a few percent for the metastable
atoms under consideration, and that 
\begin{equation}\label{C_3E_a}
  E_{\mathrm{a}}=\frac{4C_6}{3\alpha^2(0)} \mathrm , 
\end{equation}
where $C_6$ is the  {\it interatomic} vdW coefficient which is well
known for He* and Ne* from recent calculations \cite{Yan98,neon,neon2}
and gives $E_a=1.18$\,eV for He* and $E_a=2.04$\,eV for Ne*.
The static polarizability $\alpha(0)$ for He* and Ne* is given in the
standard literature 
 as $\alpha(0)=46.8$\,\AA$^3$ and $\alpha(0)=27.6 $\,\AA$^3$, respectively. For the grating
 material, $g_0$ and 
$E_{\mathrm S}$ are not known, but $g(\mathrm i\omega)$ can be 
determined via Eq.\,(\ref{g_0}) and Kramers-Kronig relations
\cite{Jackson}, once the imaginary part $\epsilon _2(\omega)$ of the
dielectric function is given. 

From optical measurements \cite{Savas01} on the
low-pressure chemical vapor deposited (LPCVD) silicon
nitride material of the grating in use, $\epsilon _2(\omega)$ has been
determined from 1\,eV to about 6\,eV. It has been shown recently that
for LPCVD SiN$_x$ $\epsilon _2(\omega)$ 
{\it over all frequencies} is essentially given by the Tauc-Lorentz
formula \cite{Zollner00}
\begin{equation}\label{TaucLorentz}
  \epsilon _2(\omega)=\Theta(\omega-\Omega_{\mathrm T})\frac{A\Omega
    \Gamma (\omega -\Omega_{\mathrm T})^2}{[(\omega ^2-\Omega
    ^2)^2+\Gamma ^2\omega ^2]\,\omega} \mathrm ,
\end{equation}
where $\Theta$ is the step function, $\hbar\Omega_{\mathrm T}$
represents the optical band gap of the material,
$A,\Omega,\Gamma$ are the strength, frequency, and spectral width,
respectively, of 
the characteristic electronic transitions within the solid. 
The $\epsilon _2(\omega)$ that results from the optical measurements
\cite{Savas01} 
is perfectly described by Eq.\,(\ref{TaucLorentz}), with
$\hbar\Omega_{\mathrm T}=2.29\,\mathrm{eV}$,  
$\hbar A=74.5\,\mathrm{eV}$, $\hbar\Omega=7.17\,\mathrm{eV}$, 
and $\hbar\Gamma=7.62\,\mathrm{eV}$. 

With the response function $g(i\omega)$ of the solid determined from
$\epsilon _2(\omega)$ as described above and using for
$\alpha(\mathrm{i}\omega)$ 
the approximate Eq.\,(\ref{alpha}), the Lifshitz formula
Eq.\,(\ref{Lifshitz}) 
yields $C_3({\rm He^*})= (3.9 \pm 0.1)$\,meV\,nm$^3$ 
and $C_3({\rm Ne^*})= (3.6 \pm 0.1)$\,meV\,nm$^3$, in agreement within
errors with the present experimental 
values $C_3$(He*)$=(4.1 \pm 1.0)$\,meV\,nm$^3$ and $C_3$(Ne*)$=(2.8\pm
1.0)$\,meV\,nm$^3$.  

Since for He* the dynamical
polarizability at imaginary frequencies $\alpha(\mathrm
i\omega)$ is known over nearly the entire frequency range from
theoretical calculations \cite{Yan98,Bishop93}, the single
oscillator approximation Eq.\,(\ref{alpha}) can be checked with the
more exact $\alpha$ in Eq.\,(\ref{Lifshitz}) which leads to 
$C_3$(He*)$=(4.0\pm0.1)$\,meV\,nm$^3$ -- only a 3\% correction,  as 
expected from Ref.\cite{Tang69}. 

Identifying in Eq.\,(\ref{g}) the oscillator strength $g_0$ with the
static limit $g(\mathrm i\omega\rightarrow 0)= 0.588$ of
Eq.\,(\ref{g}) that is
extracted from the optical data, and 
the theoretical values of $C_3$ are reproduced to within less than
10\% by the formula \cite{Vidali81}
\begin{equation}\label{Vidali}
  C_3^{\mathrm{1osc}}=\alpha(0)\,g_0\frac{E_{\mathrm a}E_{\mathrm
      S}}{8(E_{\mathrm a}+E_{\mathrm S})} \,\mathrm , 
\end{equation}
which corresponds to Eq.\,(\ref{Lifshitz}) with the one-oscillator approximations,
Eqs.\,(\ref{alpha}),(\ref{g}), if one assumes for $E_S$ a value of 13\,eV.

Table I summarizes the above results and Fig. 2 displays the
newly found values of $C_3$ for the metastable atoms together with
those measured earlier for groundstate atoms \cite{Grisenti99},
plotted versus the static atomic polarizability.

In summary, accurate atom-surface van der Waals coefficients $C_3$ for the
highly polarizable metastable excited He* $^3$S$_1$ and Ne* $^3$P$_2$
atoms have been determined for the first time, using atomic
diffraction off a silicon nitride transmission grating with a new
theoretical approach. The
experimental results are in agreement with the theory of van der
Waals forces according to Lifshitz within the experimental errors. 
The present experimental approach can
be refined to measure higher-order multipole
moments of the long-range potential between the atom and the
surface. In the future we plan to insert a grating into one of the
beams of a Mach-Zehnder-type   
interferometer in order to measure atom-surface vdW
potentials with interferometric precision. 

We are extremely grateful to Tim Savas and
Hank Smith (both MIT) for providing the SiN$_x$ grating which
has made these experiments possible. We thank L. Bruch and K. T. Tang for 
valuable discussions. This research has
been supported in part by the Deutsche Forschungsgemeinschaft.

\newpage

FIG. 1. Experimental  diffraction patterns of (a) He$^*$ at
        $v=2347\,\mathrm{m/s}$ and (b) Ne$^*$ at 
        $v=873\,\mathrm{m/s}$. The beam 
        divergence of $\Delta\vartheta=0.1\,\mathrm{mrad}$ and the small
        effective velocity spread of $\Delta v/u=3\,\%$ allow ten
        principal maxima to be recorded. The background signal in both
        cases is about 10 counts/s. Solid lines have been added to
        guide the eye.
      
\vspace{1cm}

   FIG. 2. Comparison of  measured ($\times$) and  theoretical values
      for $C_3$. $\bullet$:  Eq.\,(\ref{Lifshitz}) with $\alpha$ from
      Eqs.\,(\ref{alpha}) and (\ref{C_3E_a}); 
      {$\Diamond$}: Eq.\,(\ref{Lifshitz}) with $\alpha$ exact [23].
      The data points on the straight line (Hoinkes
      approximation [20]) are for groundstate particles
      [3], namely, with increasing $\alpha$: He, Ne, D$_2$, Ar, Kr.

\vspace{3cm}

\begin{table}\caption{Experimental results for $C_3$ (in
    $\mathrm{meV\,nm^3}$) for He* and Ne* compared to  
    different theoretical expressions.}
  \begin{center}
    \begin{tabular}{lll}
      $C_3(\mathrm{He^*})$   & $C_3(\mathrm{Ne^*})$ & \\ 
      \hline
      $4.1\pm1.0$ & $2.8\pm1.0$ & Experiment\\
      $3.9\pm 0.1$ & $3.6\pm0.1$ & Theory: $\alpha$ from
      Eqs. (\ref{alpha}), (\ref{C_3E_a})\\ 
      $4.0\pm0.1$ & ~~~~~-   & Theory: $\alpha$ from Ref. \cite{Yan98}\\
          \end{tabular}
  \end{center}
\end{table} 

\begin{thebibliography}{abcdefghijkl99}
\bibitem{Bruch97}
  L.~W.~Bruch, M.~W.~Cole, and E.~Zaremba,
  {\it Physical Adsorption: Forces and Phenomena} 
  (Clarendon, Oxford, 1997)
\bibitem{Lifshitz56}  E.~M.~Lifshitz, JETP {\bf 2},1 73 (1956);
  I.~E.~Dzyaloshinskii, E.~M.~Lifshitz, and L.~P.~Pitaevskii,
  Adv.~Phys. {\bf 10}, 165 (1961)
\bibitem{Grisenti99}
  R.~E. Grisenti, W.~Sch\"ollkopf, J.~P.~Toennies,
  G.~C.~Hegerfeldt, and T.~K\"ohler, 
  Phys.~Rev.~Lett. {\bf 83}, 1755 (1999)
\bibitem{Anderson88}
  A.~Anderson, S.~Haroche, E.~A.~Hinds, W.~Jhe, and D.~Meschede,
  Phys.~Rev.~A {\bf 37}, 3594 (1988)
\bibitem{Jhe87}
  W.~Jhe, A.~Anderson, E.~A.~Hinds, D.~Meschede, L.~Moi, and
  S.~Haroche, Phys.~Rev.~Lett. {\bf 58}, 666 (1987);  V.~Sandoghar,
  C.~I.~Sukenik, E.~A.~Hinds, and S.~Haroche,  
  Phys.~Rev.~Lett. {\bf 68}, 3432 (1992).
\bibitem{Failache99}
  H.~Failache, S.~Saltiel,~M.~Fichet, D.~Bloch, and M.~Ducloy,
  Phys.~Rev.~Lett {\bf 83}, 5467 (1999)
\bibitem{Boustimi01}
  M.~Boustimi, B.~Viaris de Lesegno, J.~Baudon, J.~Robert, and M.~Ducloy,
  Phys.~Rev.~Lett. {\bf 86}, 2766 (2001)
\bibitem{Shimizu01}
  F.~Shimizu, Phys.~Rev.~Lett. {\bf 86}, 987 (2001); F. Shimizu and
  J. Fujita, Phys. Rev. Lett. {\bf 88}, 123201 (2002)
\bibitem{Adams94}
  C.~S.~Adams, M.~Sigel, and J.~Mlynek, Phys.~Rep. {\bf 240}, 143
  (1994); 
  O.~Carnal, A.~Faulstich, and J.~Mlynek, Appl.~Phys.~B
  {\bf 53}, 88 (1991) 
\bibitem{surf}
  Y.~Harada, S.~Masuda, and H.~Ozaki, Chem.~Rev.~{\bf 97}, 1897
  (1997); H.\,Hotop, Exp.\,Meth.\,Phys.\,Sci. {\bf 29B}, 191 (1996)
\bibitem{Robert01}
  A.~Robert, O.~Sirjean, A.~Browaeys, J.~Poupard, S.~Nowak, D.~Boiron,
  C.~I.~Westbrook, and A.~Aspect, Science {\bf 292}, 461 (2001); 
  F.~Pereira~Dos~Santos, J.~L\'eonard, Junmin~Wang, C.~J.~Barrelet,
  F.~Perales, E.~Rasel, C.~S.~Unnikrishnan, M.~Leduc, and
  C.~Cohen-Tannoudji, Phys.~Rev.~Lett. {\bf 86}, 3459 (2001) 
\bibitem{Sengstock01} P.~Engels, W.~Ertmer, and K.~Sengstock,
  Optics Commun. {\bf }204, 185 (2002)
\bibitem{Ste83}C.~Fabre and S.~Haroche in: {\it Rydberg States of
    Atoms and Molecules}, 
  R.~F.~Stebbings and F.~B.~Dunning, eds. (Cambridge University Press, 1983).
\bibitem{Hutson86}
  J.~M.~Hutson, P.~W.~Fowler and E.~Zaremba, Surf.~Sci. {\bf 175},
  L775 (1986)
\bibitem{Zeilinger2002}E.g., to B. Brezger, L. Hackerm\"uller,
  S. Uttenthaler, J. Petschinka, M. Arndt, and A. Zeilinger,
  Phys. Rev. Lett. {\bf 88}, 100404 (2002).
\bibitem{Grisenti00}
  R.~E.~Grisenti, W.~Sch\"ollkopf, J.~P.~Toennies, J.~R.~Manson,
  T.~A.~Savas, and H.~I.~Smith, Phys.~Rev.~A {\bf 61},
  033608 (2000) 
\bibitem{Fouquet98}
  P. Fouquet, P.~K.~Day, and G.~Witte, Surf.~Sci.~{\bf 400}, 140 (1998)
\bibitem{Siska93}
  P.~E.~Siska, Rev.~Mod.~Phys. {\bf 65}, 337 (1993)
\bibitem{Joachain83}
  C.~J.~Joachain, {\it Quantum Collision Theory}, 3rd Ed. (North
  Holland, 1983)
\bibitem{Hoinkes80}
  H.~Hoinkes, Rev.~Mod.~Phys. {\bf 52}, 933 (1980)
\bibitem{Vidali81}
  G.~Vidali and M.~W.~Cole, Surf.~Sci. {\bf 110}, 10 (1981)
\bibitem{Tang69} K. T. Tang, Phys. Rev. {\bf 177}, 108 (1969)
\bibitem{Yan98}
  Z.-C.~Yan and J.~F.~Babb, Phys.~Rev.~A {\bf 58}, 1247 (1998)
\bibitem{neon} 
  A.~Derevianko and A.~Dalgarno, Phys.~Rev.~A {\bf 62}, 062501
  (2000)
\bibitem{neon2}
  S.~Kotochigova,  E.~Tiesinga, and I.~Tupitsyn,
  Phys. Rev.~A {\bf 61}, 042712 (2000); 
  M.~R.~Doery, E.~J.~D.~Vredenbregt, S.~S.~Op de Beek,
  H.~C.~Beijerinck, and B.~J.~Verhaar, Phys.~Rev.~A {\bf 58}, 3673
  (1998)
\bibitem{Jackson}J. D. Jackson, {\it Classical Electrodynamics}, 3rd
  Edition (Wiley, New York, 1999)
\bibitem{Savas01}
  T.~A.~Savas, private communication to J.~P.~T.
\bibitem{Zollner00}
  S.~Zollner, E.~Apen, AIP Conference Proceedings {\bf 550},
  D.~G.~Seiler {\it et al.} eds., 2001, p.~532 
\bibitem{Bishop93}
  D.~M.~Bishop and J.~Pipin, Int.~J.~Quant.~Chem. {\bf 47}, 129 (1993);
  R.~M.~Glover and F.~Weinhold, J.~Chem.~Phys. {\bf 66}, 191 (1977)
\end{thebibliography}
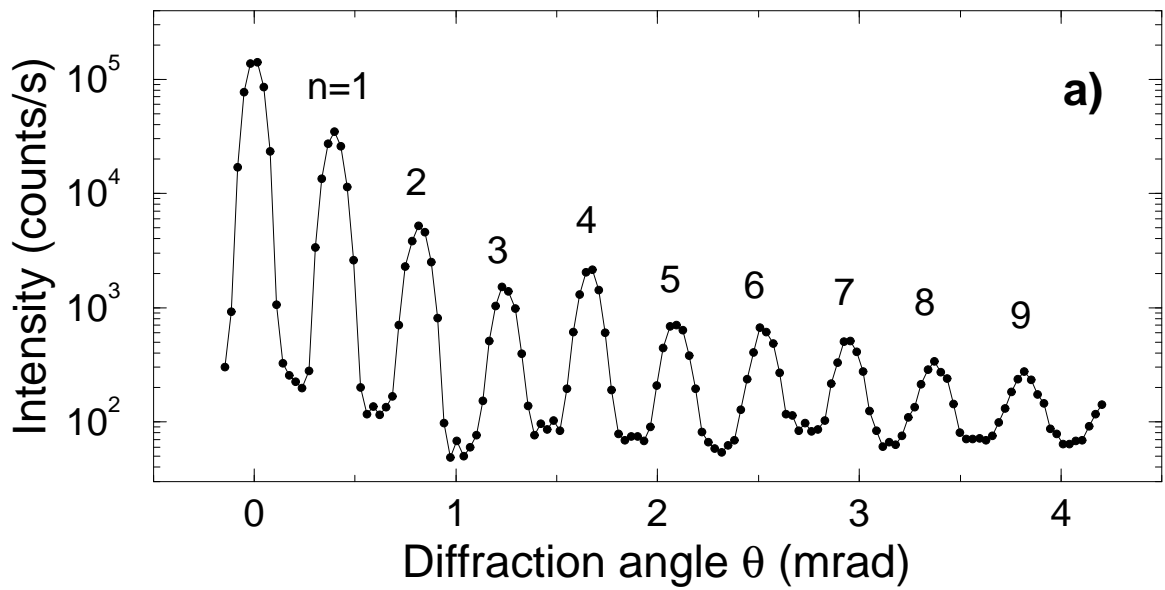
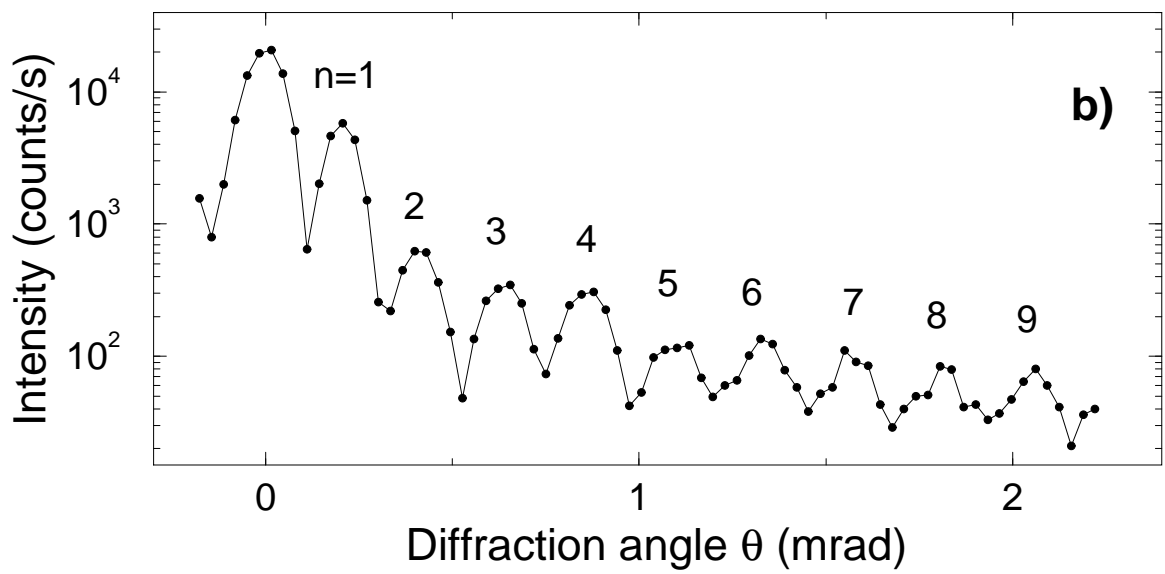
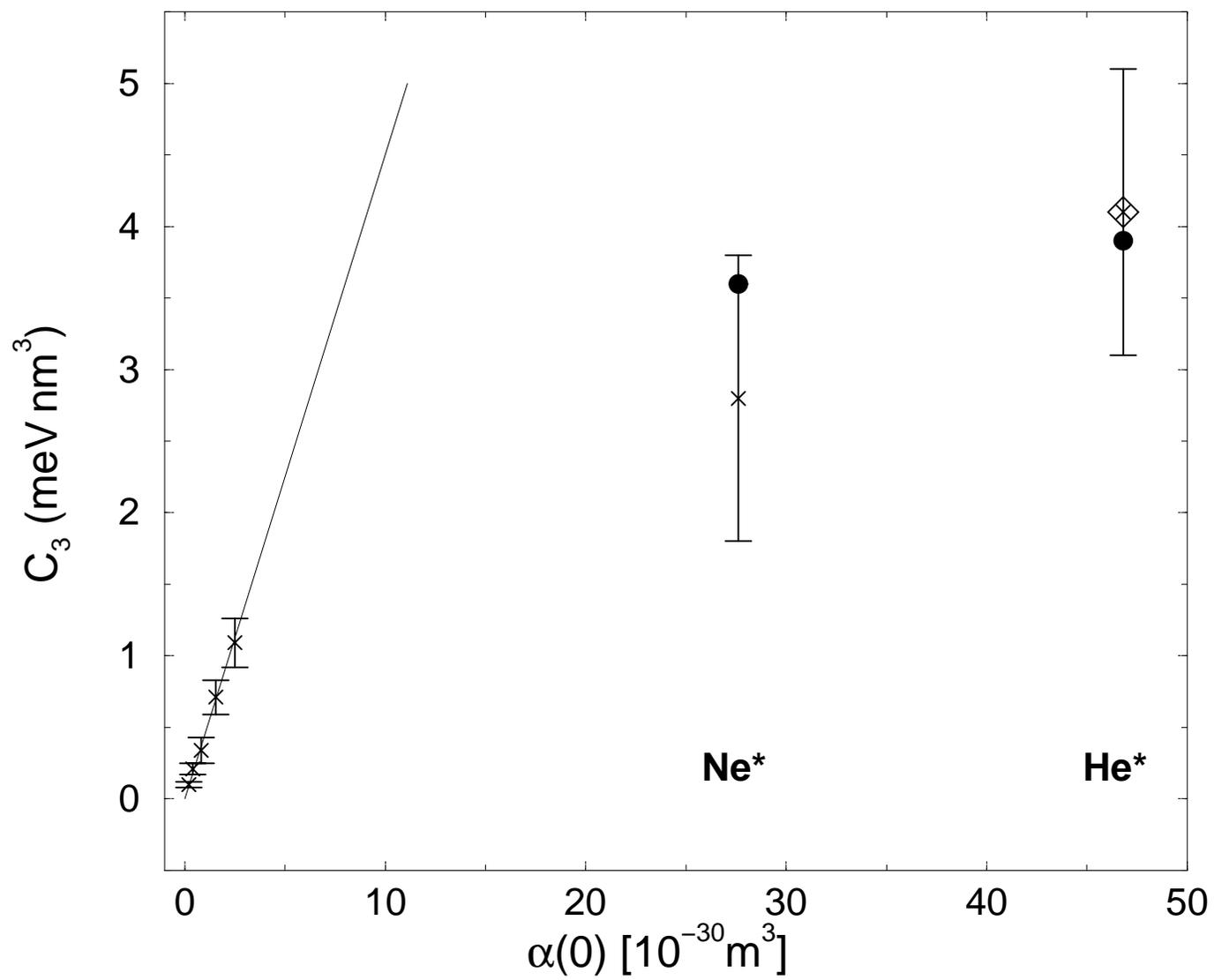
\end{document}